\documentclass[]{emulateapj}
\usepackage{amsmath}
\usepackage{amssymb}
\usepackage{amstext}
\usepackage{apjfonts}
\usepackage{graphicx}
\usepackage{epsfig}

\usepackage{color}
\definecolor{myColor}{rgb}{0.9,0.9,0.9}

\begin{document}

\shorttitle{IMBH-ULXs}
\shortauthors{Madhusudhan, et al.}
\title{MODELS OF ULTRALUMINOUS X-RAY SOURCES WITH INTERMEDIATE-MASS BLACK HOLES}

\author{N. Madhusudhan\altaffilmark{1},  S. Justham\altaffilmark{2}, L. Nelson\altaffilmark{3}, B. Paxton\altaffilmark{4}, E. Pfahl\altaffilmark{4}, Ph. Podsiadlowski\altaffilmark{2}, \& S. Rappaport\altaffilmark{1} }   

\altaffiltext{1}{Department of Physics and Kavli Institute for Astrophysics and Space Research, MIT, Cambridge, MA 02139; {\tt nmadhu@mit.edu}}
\altaffiltext{2}{Department of Astrophysics, Oxford University, Oxford OX1 3RH, UK}
\altaffiltext{3}{Physics Department, Bishop's University, Lennoxville, QC Canada J1M 1Z7}
\altaffiltext{4}{KITP, University of California at Santa Barbara}


\begin{abstract}

We have computed models for ultraluminous X-ray sources (``ULXs'') consisting of a black-hole accretor of intermediate mass (``IMBH''; e.g., $\sim$$1000~M_\odot$) and a captured donor star.  For each of four different sets of initial donor masses and orbital separations, we computed 30,000 binary evolution models using a full Henyey stellar evolution code.  To our knowledge this is the first time that a population of X-ray binaries this large has been carried out with other than approximation methods, and it serves to demonstrate the feasibility of this approach to large-scale population studies of mass-transfer binaries.  In the present study, we find that in order to have a plausible efficiency for producing active ULX systems with IMBHs having luminosities $\gtrsim 10^{40}$ ergs s$^{-1}$, there are two basic requirements for the capture of companion/donor stars.  First, the donor stars should be massive, i.e., $\gtrsim 8~M_\odot$.   Second, the initial orbital separations, after circularization, should be close, i.e., $\lesssim 6-30$ times the radius of the donor star when on the main sequence.   Even under these optimistic conditions, we show that the production rate of IMBH-ULX systems may fall short of the observed values by factors of 10-100.

\end{abstract}


\section{Introduction}
\label{sec:intro}

Ultraluminous X-ray sources (ULXs) are off-nucleus sources in external 
galaxies with luminosities of $10^{39} \lesssim L_x \lesssim 10^{41}$ 
ergs s$^{-1}$. They have been discovered in great numbers with {\em ROSAT}, 
{\em Chandra}, and {\em XMM-Newton} (Fabbiano 1989; Roberts \& 
Warwick 2000; Ptak \& Colbert 2004; 
Fabbiano \& White 2004; Colbert \& Miller 2004).  For reference, the 
Eddington limit of a black hole accretor of $\sim$$10~M_\odot$ is in the range 
of $\sim$$1.5-3 \times 10^{39}$ ergs s$^{-1}$, depending on the chemical 
composition of the accreted material.  ULXs are especially prevalent in 
galaxies with starburst activity, including ones that have likely undergone a 
recent dynamical encounter (e.g., Fabbiano, Zezas, \& Murray 2001; Wolter \& Trinchieri
2004; Fabbiano \& White 2004; Colbert \&
Miller 2004). Many, but not all, of the ULXs are identified with 
young star-forming regions (see, e.g., Fabbiano \& White 2004). 

The central question regarding
this important class of sources is whether they represent an extension
in the luminosity function of binary X-ray sources containing neutron
stars and stellar-mass black holes (BHs), or a new class of objects,
e.g., systems containing intermediate-mass black holes ($10^2-10^4
~M_\odot$; hereafter ``IMBHs''). 

Colbert \& Mushotzky (1999) first suggested that some ULXs harbor 
IMBHs. The motivation for this is clear. 
The Eddington limit for a $1000~M_\odot$ black hole is $\sim 10^{41}$ ergs 
s$^{-1}$, compared to only $\sim$$10^{38}$ ergs s$^{-1}$ for a typical neutron
star, and $\sim$10 times this value for a stellar-mass black hole accretor.  
Moreover, the spectra from IMBHs might be expected to have low inner-disk
temperatures, as is inferred for some of the ULXs (Miller et al.\
2003; Miller et al. 2004; Cropper et al. 2004).  

A number of ideas have been put forth for ways to circumvent the
problem of how $\sim$$10 \,M_\odot$ BHs could have apparent $L_x$
values as high as $10^{40}-10^{41}$ ergs s$^{-1}$.  King et al.\
(2001) suggested that the radiation may be geometrically beamed 
by a thick accretion disk so that the true value of $L_x$ does not, in fact, 
exceed the Eddington limit.  K\"ording, Falcke, \& Markoff (2002) proposed that the
apparently super-Eddington ULXs are actually emission from microblazar
jets that are relativistically beamed along our line of sight.  However, studies 
of the giant ionization nebulae surrounding a number of the ULXs (Pakull \&
Mirioni 2003) seem to confirm the full luminosity inferred from the
X-ray measurements.  Begelman (2002) and Ruszkowski \& Begelman (2003)
found that in radiation pressure dominated accretion disks
super-Eddington accretion rates of a factor of $\sim$10 can be
achieved due to the existence of a photon-bubble instability in
magnetically constrained plasmas.  

In two previous studies we have explored the evolution of binary stars consisting 
of $2-17$ $M_\odot$ donor stars and {\em stellar-mass} BHs as possible model systems 
for ULX sources (Podsiadlowski, Rappaport, \& Han 2003; Rappaport, Podsiadlowski, \& 
Pfahl 2005a).  These BH binaries are assumed to form from massive primordial 
binaries in regions of intense star formation.  In our earlier work we showed that the 
formation efficiencies for such systems were adequate to explain the observed 
formation rates of ULX systems.  However, the Eddington limit would somehow 
have to be violated by a factor of $\sim$$10-30$, and even then, the mass transfer 
rates were only marginally sufficient to explain the luminosities of the most powerful ULXs.  

In this work we investigate the evolution of a population of binaries containing 
an IMBH with a representative mass of $1000~M_\odot$.  The mechanisms by 
which an IMBH is created in star-forming regions and acquires a companion/donor 
star are highly uncertain.  One proposed scenario for their formation in star clusters
involves a runaway stellar collision process leading to the production of a very 
massive star which somehow evolves to core collapse before much of its 
envelope has been lost.  While the dynamics of the runaway collision may be 
understood quantitatively (e.g., Portegies Zwart \& McMillan 2002; Portegies 
Zwart, Dewi, \& Maccarone 2004; G\"urkan, Freitag, \& Rasio 2004) there is a 
great deal of uncertainty over how, and whether, 
such a massive star evolves to core collapse and the formation of an IMBH. 
If we allow for the presence of an IMBH at the center of a star cluster shortly
after the formation of the cluster itself, then one must consider how the
IMBH acquires a mass-transferring companion star.  Presumably the IMBH
could capture a companion from stars near the cluster core 
via either tidal capture or 
exchange encounters with primordial binaries in the cluster (Hopman, 
Portegies Zwart, \& Alexander 2004; Pfahl 2005).  After the IMBH has 
acquired a companion, but before steady mass transfer is possible, the star 
must survive tidal circularization and, more specifically, the tidal heating that 
accompanies the process.  There has been much discussion in the literature 
about this issue (e.g., Ray, Kembhavi, \& Antia 1987; Podsiadlowski 1996; 
Alexander \& Morris 2003).  Finally,
the binary system has to survive encounters with other passing stars---either 
single or binary stars.  For earlier work on IMBH-ULX binary evolution see, e.g.,
Rappaport, Podsiadlowski, \& Pfahl (2005b), Kalogera et al. (2005), and 
Patruno et al. (2005).  For a very recent investigation of the dynamical capture 
of companion/donor stars by an IMBH, see Blecha et al. (2005).

Blecha et al. (2005) found a wide range of semimajor axes for the companion stars 
captured by the IMBH, with the majority in the range $\sim 100-10^4\,R_\odot$
(and $\sim$90\% lying between 30 and $3 \times 10^4$ $R_\odot$).  We know 
from our previous work, and will show in the present study, that very wide binary 
systems do not produce ULXs with interesting durations during the lifetime of the 
star cluster.  Therefore, we have chosen to study systems with initial orbital separations in the 
range of $\sim 20-1000 \, R_\odot$ wherein $\sim$40\% of the Blecha et al. (2005) 
systems are found.  We further assume that the IMBH has only a single stellar 
companion which is counter to the main findings of Blecha et al. (2005).  However,
the assumption of a single companion would be quite reasonable for Model C  (low 
stellar density) of Blecha et al. (2005) and for times up to $\sim$30 Myr, especially
when taking into account that these authors assumed a primordial binary fraction of 100\%.

In this work we follow 30,000 binary evolution sequences for each of four different distributions 
of starting masses and orbital periods.  The close orbits are assumed to have 
circularized by tidal friction.  Each binary is followed with a full Henyey 
stellar evolution code until the entire envelope of the donor star has been transferred.  
To the best of our knowledge, this is the first time that a binary population 
study this large has been carried out with a full Henyey code.  \footnote{Prior to this, the largest number of mass-transfer binaries computed in a single study (in the published record) was 5500 (Nelson \& Eggleton 2001).}

\section{Population Study}
\label{sec:obs}

The stellar evolution of the donor stars, including mass loss, was followed with 
{\tt EZ} which is a stripped down, rewritten version of a subset of the stellar evolution code developed by P. P. Eggleton (Paxton 2004).  The physics of the program is unchanged 
from Eggleton's (essentially as described in Pols et al.\,1995), but the structure of the 
code has been modified to facilitate experiments involving programmed control 
of parameters.  There are zero-age main-sequence (ZAMS) starting models for a 
variety of metallicities (from $Z=10^{-4}$ to $Z=0.03$) and masses 
(from 0.1 to 100 $M_\odot$), with arbitrary 
starting masses created by interpolation.  A user-provided procedure is called 
between~steps of the evolution to inspect the current state, to make changes in 
parameters, and to decide when and what to record to log files.  
The source code and data for {\tt EZ} can be downloaded from 
the web at $<$http://theory.kitp.ucsb.edu/$\sim$paxton$>$.  For all models in this particular study, the number of stellar mesh-points was fixed at 200 in the interest of minimizing computation time.

\begin{deluxetable}{c c c c c}
\tablewidth{0pt}
\tablecaption{ULX-IMBH Population Models
\label{tab:opt}}
\tablehead{
\colhead{Model} & \colhead{$M_{\rm low}$\tablenotemark{a}} &
\colhead{$M_{\rm high}$\tablenotemark{b}} & \colhead{$f_{\rm max}$\tablenotemark{c}} & \colhead{$a_{\rm min}-a_{\rm max}$\tablenotemark{d}} }
\startdata
A & 5 & 50 & 15 & $40-900$ \\
B & 1 & 10 & 15 & $20-600$ \\
C & 5 & 50 & 3 & $40-200$ \\
D & 1 & 10 & 3 & $20-130$ \\
\enddata
\tablenotetext{a}{Lower limit on the donor mass, in units of $M_\odot$}
\tablenotetext{b}{Upper limit on the donor mass, in units of $M_\odot$}
\tablenotetext{c}{Maximun initial orbital separation in units of that required for the donor star to fill its Roche lobe; $f \equiv a_{\rm init}/a_{\rm RL}$}
\tablenotetext{d}{Range of initial orbital separations, in units of $R_\odot$}
\end{deluxetable}

A binary ``driver'' which computes the changes in orbital separation and the mass 
loss and transfer rate (according to the appropriate conservations laws) was written 
specifically for this project.  However, it is similar to the 
procedures and approaches which we have used in our stellar evolution programs 
in the past (see, e.g., Podsiadlowski, Rappaport, \& Pfahl 2002).  In particular, we 
utilize a semi-explicit scheme for determining the mass loss at each time step based on the
instantaneous location of the Roche lobe within the atmosphere of the donor star, which is
approximated as having an exponential density distribution.  

For each binary we choose, via Monte Carlo means, the initial donor mass uniformly between $M_{\rm low}$ and $M_{\rm high}$ (see Table 1 for specific model values).  The initial orbital separation was chosen uniformly between 1 and $f_{\rm max}$ times the separation required for a ZAMS star (of the same mass) to fill its Roche lobe (see Table 1).  For Models B and D the donor masses ranged between 1 and $10 \, M_\odot$, while Models A and C featured much higher mass stars ($5-50\,M_\odot$).  Models A and B had initial orbital separations up to $f_{\rm max} = 15$, which crudely represent what might be expected from the tighter exchange encounters with passing primordial binaries (see also Blecha et al. 2005).  In contrast, Models C and D had $f_{\rm max} = 3$ to roughly approximate the case of direct tidal capture, and circularization, of a single field star by the IMBH (see, e.g., Hopman et al. 2004).  Parameters for the four illustrative distributions we have chosen to explore are summarized in Table 1.  We ran 30,000 binary evolution sequences for each of the models listed in Table 1.

\begin{figure*}
\begin{center}
\includegraphics[width=0.47\textwidth]{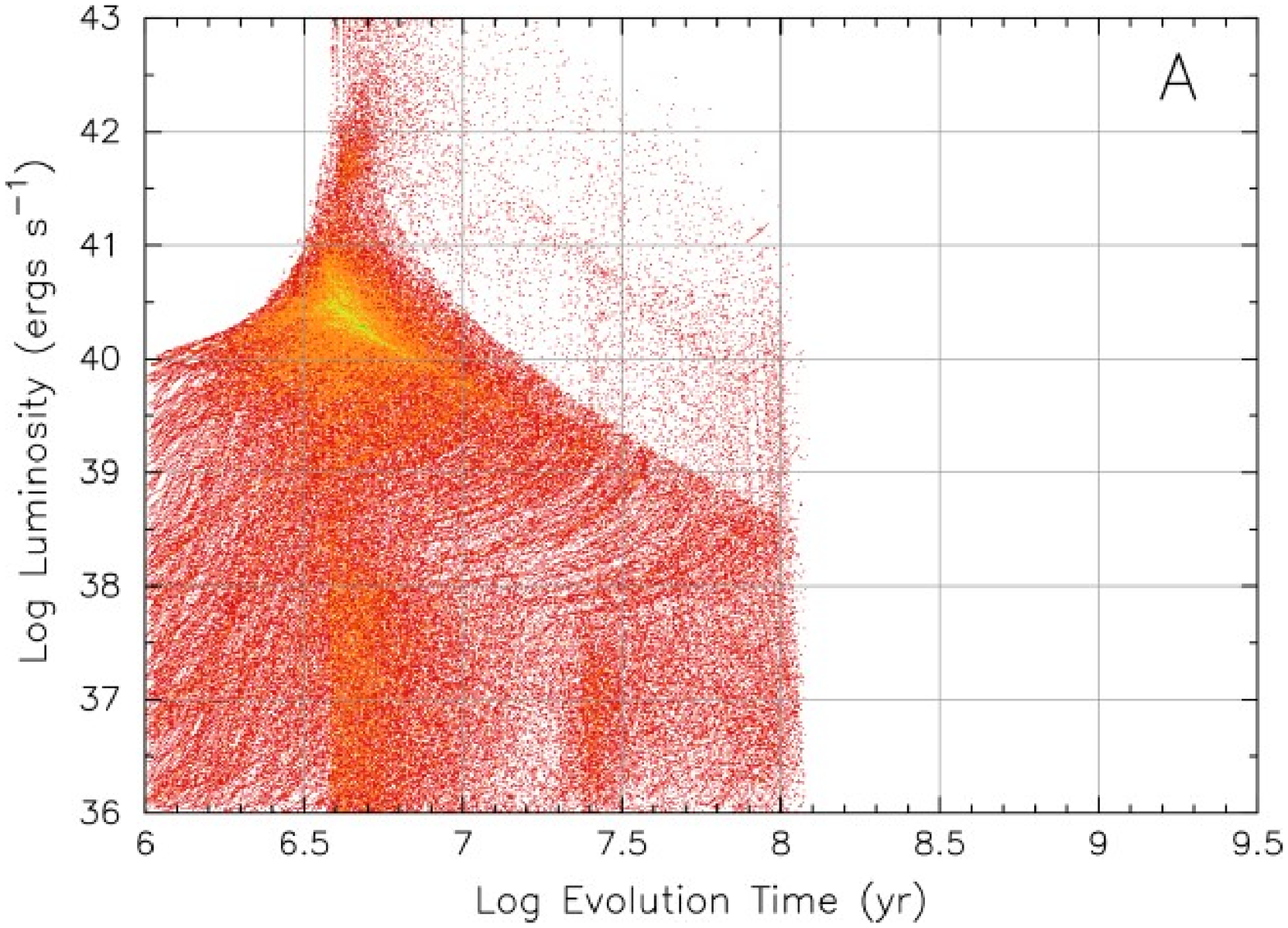}\hglue1cm
\includegraphics[width=0.47\textwidth]{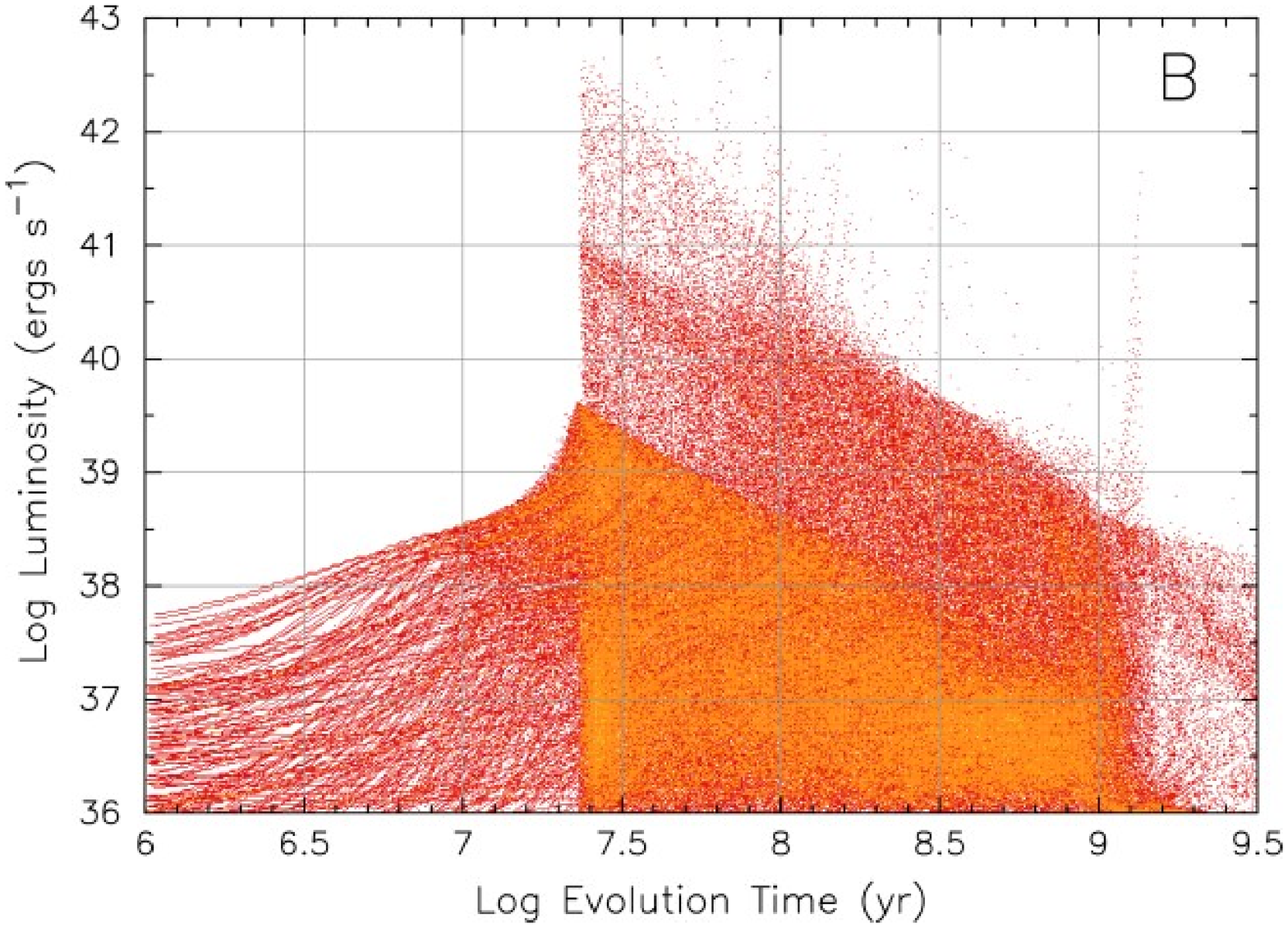}
\vglue0.2cm
\includegraphics[width=0.47\textwidth]{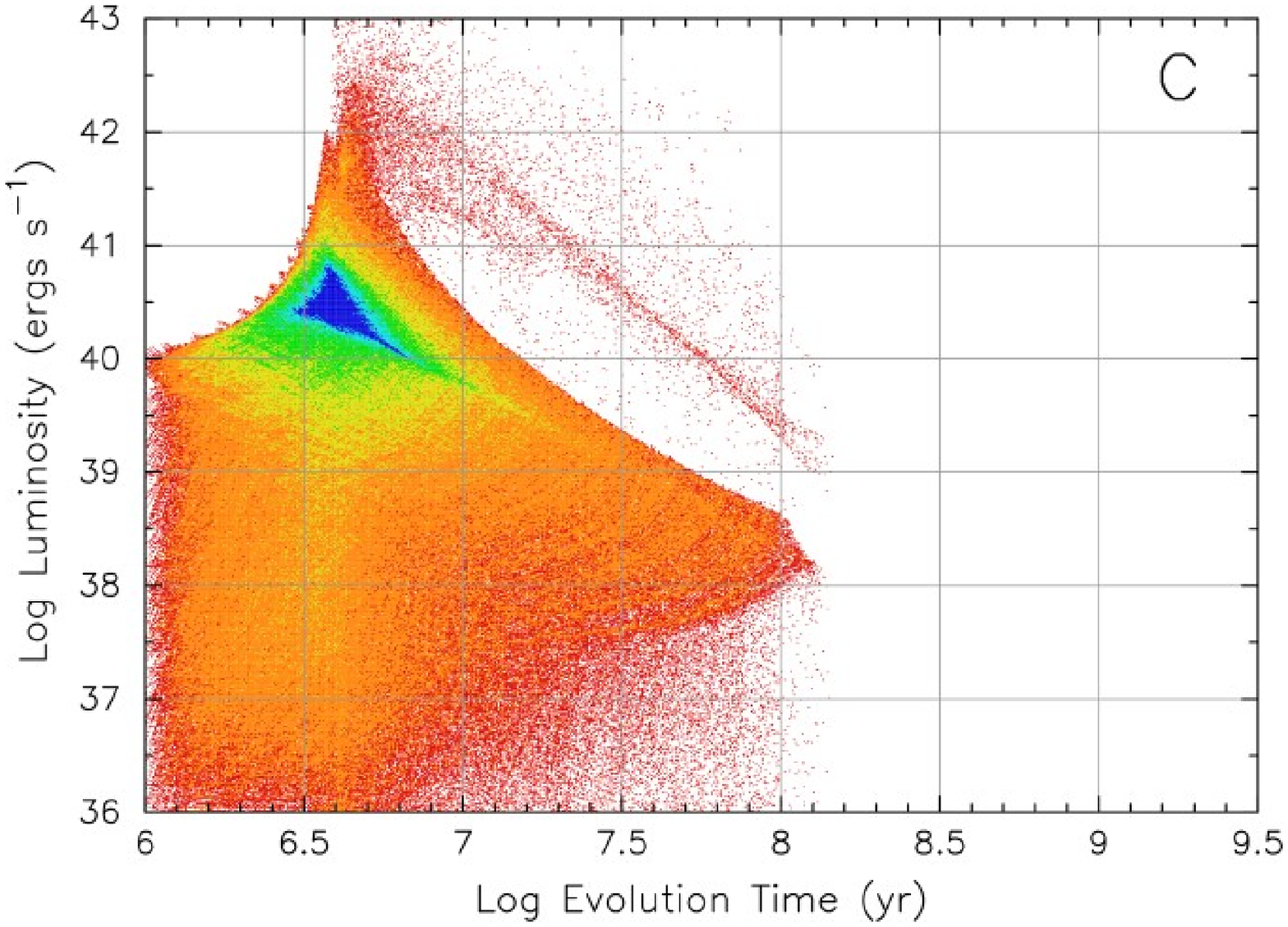}\hglue1cm
\includegraphics[width=0.47\textwidth]{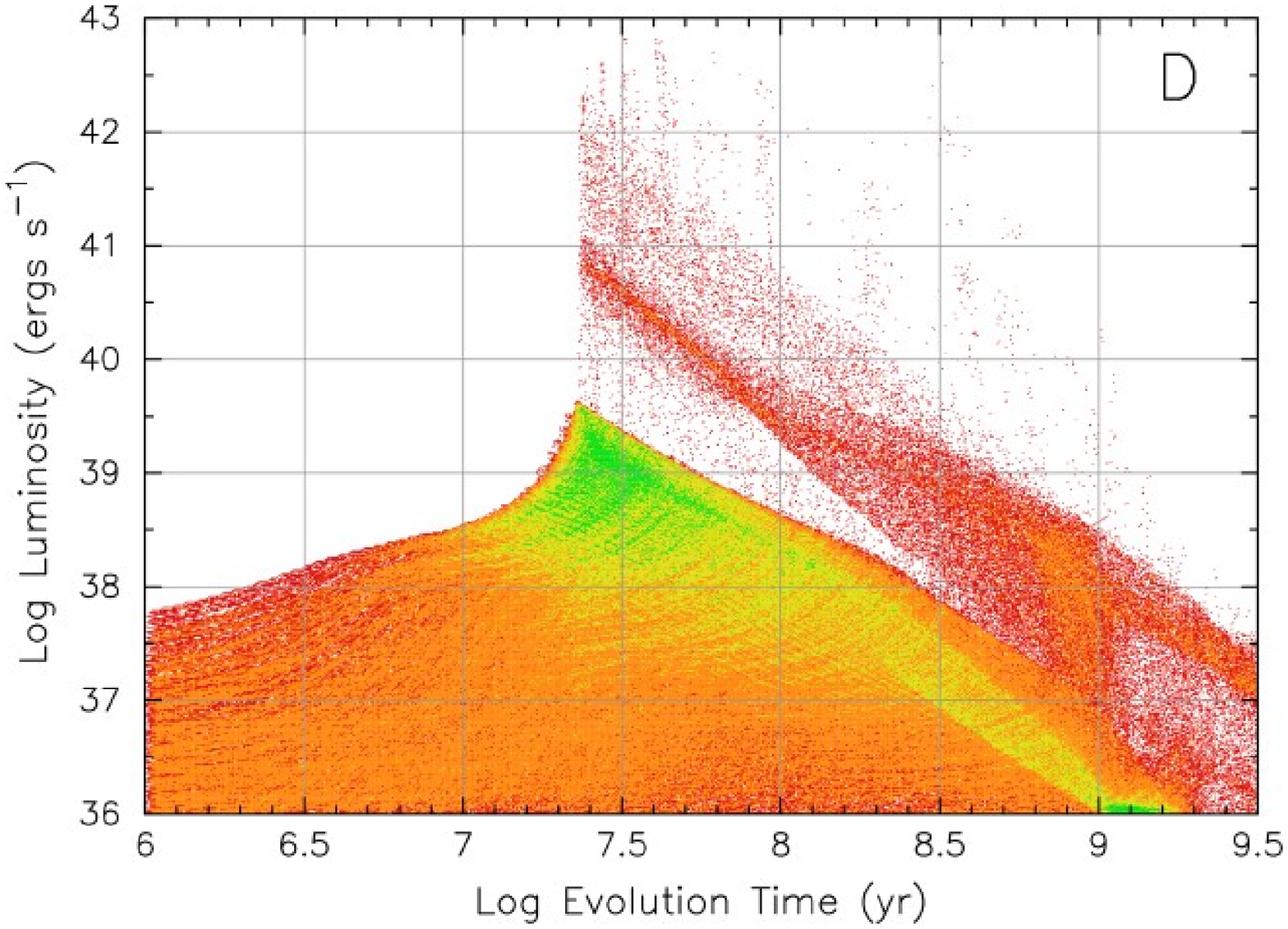}
\caption{Simulated evolution of the X-ray luminosity function of IMBH black
hole binaries with time since the birth of the host star cluster.
For each panel, evolution tracks from $30,000$ X-ray 
binaries were computed and then registered in each of the $700 \times 700$ 
pixels that are traversed. In the calculations used to produce these plots, the 
Eddington limit (at $\sim 2 \times 10^{41}$ ergs s$^{-1}$) {\em was} enforced; however, 
for clarity the $L_x$ values displayed here are without any Eddington limit.  This affects 
only a relatively small fraction of the highest luminosity systems.  The colors represent 
the square root of the relative populations, with red through red corresponding 
to actual ratios of $\sim$100 to 1.  The distributions of initial masses and orbital separations 
for the four models are specified in Table 1.}
\end{center}
\end{figure*}

All models were run using 60 nodes of the {\tt elix3} Beowulf cluster located at the University of Sherbrooke, Quebec.  The run time for each of the 4 models was $\sim$30 hours.
 
\begin{figure}[h]
\begin{center}
\includegraphics[width=0.47\textwidth]{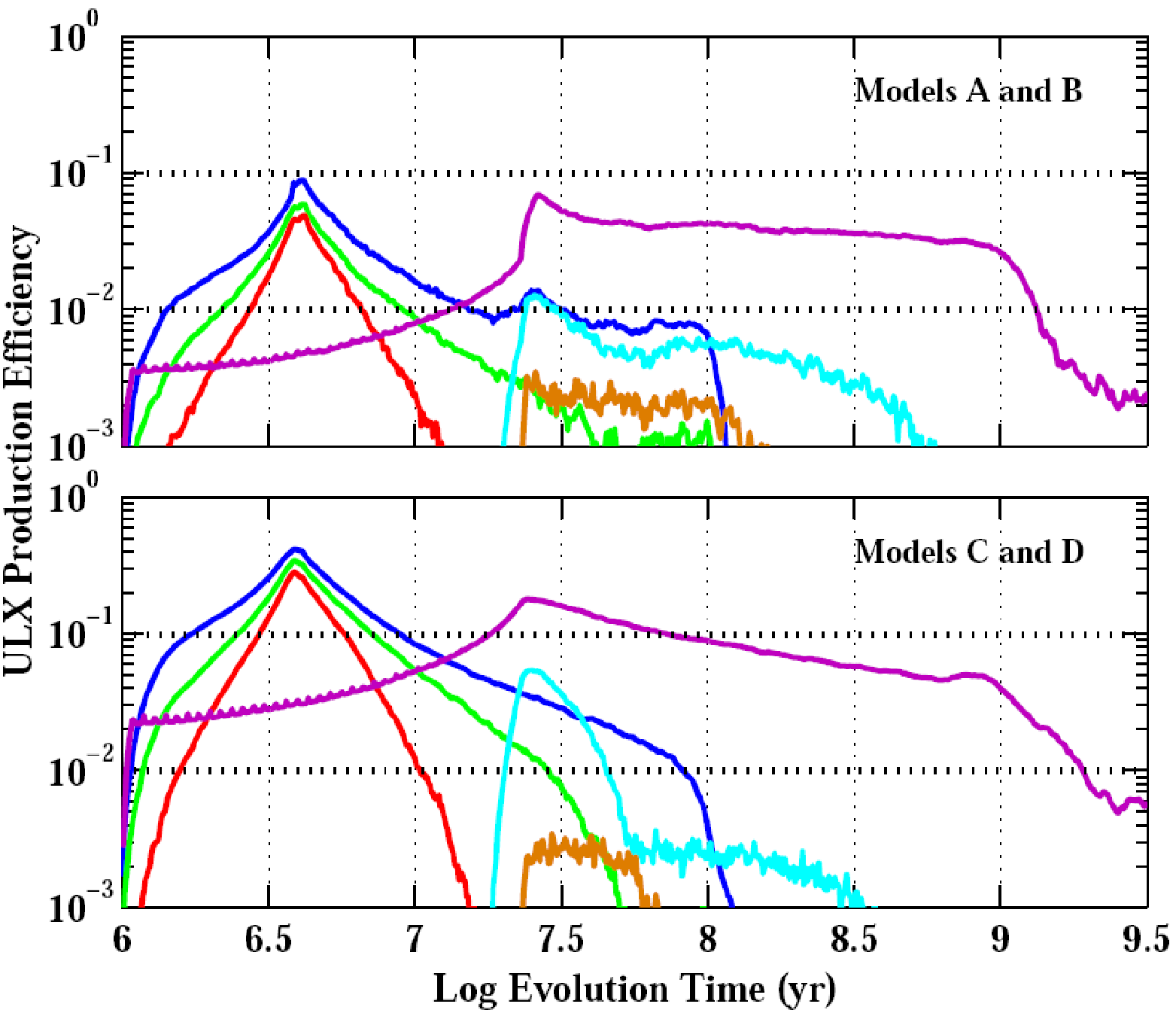}
\caption{The numbers of IMBH binaries as a function of time for three different lower 
limits on the X-ray luminosity.  Blue and purple curves -- all systems.  Green and cyan curves -- $L_x > 10^{39}$ ergs sec$^{-1}$. Red and orange curves -- $L_x > 10^{40}$ ergs sec$^{-1}$  Values are derived from 
the results in Fig. 1. Color scheme: Models A and C (blue--green--red); 
Models B and D (purple--cyan--orange).
In each case, the number of active systems at any given time, is divided by 30,000 
(the number of binaries evolved) to provide an estimate of the fractional ``efficiency'' 
for producing IMBH-ULX binaries.  
}
\end{center}
\end{figure}

\section{IMBH--ULX Results}
\label{sec:discuss}

For each evolution step of each IMBH-ULX binary, the X-ray luminosity is computed 
according to $L_x = \eta \dot M c^2$, where $\dot M$ is the mass transfer rate, 
and $\eta$ is the BH conversion efficiency 
from rest mass to radiant energy, computed according to the spin of the black-hole.  All 
IMBHs are assumed to start with zero angular momentum.  We then follow the spin history
of the IMBH (according to the formalism of Bardeen 1970); however, because of the
extreme mass ratio between the IMBH and the donor star, the dimensionless spin 
parameter of the IMBH cannot exceed $\sim$0.07, and the corresponding value of
$\eta$ does not grow appreciably above that expected for a Schwarzschild BH ($1-\sqrt{8}/3$).

For each evolution step of each IMBH-ULX binary, the X-ray luminosity and corresponding evolution time are recorded in a $700 \times 700$ matrix, by simply adding ``1" when the track crosses a particular array element.  The contributions from all of the tracks are added together.  The matrix covers 7 decades in $L_x$ and 3.5 decades in time, in equally spaced logarithmic intervals.  The results are shown in Fig.\,1 as color images with a square-root scaling in intensity to enhance the dynamic range.  Time zero is arbitrarily taken to be a common formation epoch for all stars in the cluster, as well as the capture time of a companion by the IMBH.

Figure 2 shows more quantitative plots of the results for the models ``imaged'' in Fig.\,1.  The curves in each plot show the numbers of IMBH systems that are transferring mass at any given time -- after their formation -- with values of $L_x$ that exceed $10^{36},~10^{39}, ~{\rm and}~10^{40}$ ergs s$^{-1}$.  The initial orbital periods of the systems being computed are typically $\sim 1\rightarrow f_{\rm max}^{3/2}$ days, and end with $P_{\rm orb}$ often exceeding a year.  Note that even for Models A and C, which emphasize massive donor stars, a maximum of $\sim$4\% and $\sim$25\%, respectively, of systems have $L_x \gtrsim 10^{40}$ ergs s$^{-1}$, but only for a brief interval of $\sim$1 Myr during the star cluster's active lifetime.

From Figs.\,1 and 2 we see that Models A and C, with higher-mass donor stars, produce generally higher $L_x$ values than Models B and D, and the systems are most ``active" between 3 and 30 Myr.  By contrast, Models B and D, with lower-mass donors, typically yield significantly lower values of $L_x$, but these systems appear over a much longer range of timescales, i.e., out to $\sim 10^9$ yr.  For the latter models, there are ULX-luminosity sources between $\sim$50--500 Myr due to donor stars that are ascending the giant branch; however, these are probably too rare (fewer than $\sim$0.003 of all the systems) to explain many ULXs. These differences between systems with higher and lower mass donor stars can be explained by the fact that higher mass stars have shorter nuclear evolution timescales.  This leads to shorter overall lifetimes of the donors and correspondingly higher values of $\dot M$.  Systems on the giant branch, in all four panels, but especially Models B and D, can be recognized as a red band above the main body of systems in which the donor star is still on the main sequence.  The vertical edge at $\sim$23 Myr in Model B (Fig.\,1) marks the evolution time required for a 10 $M_\odot$ star (the most massive in the model) to ascend the giant branch.  Due to the wide initial separations, most of the systems in Model B do not commence mass transfer until the donor evolves well beyond the main sequence.

The distribution of initial orbital separations is equally important in determining the production of ULXs as is the donor mass.  Models C and D, with closer initial orbital separations, have generally significantly larger populations of active ULXs than for their counterpart Models A and B.  This is true since in initially wide systems -- by the time mass transfer commences -- the donor star is already quite evolved, and the remaining evolution time on the giant branch is rather short.  Hence, these systems do not persist as ULXs for very long, and there is a correspondingly small probability of finding the system in an active state.

Finally, we comment on a technical issue of the binary evolutions.  We find that for some binary models the donor star envelope is not completely transferred to the IMBH before the stellar code terminates.  This occurs predominantly for the most massive donor stars (i.e., initial masses $\gtrsim 40\,M_\odot$) and wide initial orbital separations (i.e., $f \gtrsim 7$, where $f$ is defined in Table 1); thus, it affects mostly Model A.  The computational problems begin to occur close to the point of core He ignition when the rapid readjustment of the internal stellar structure, coupled with high mass transfer rates, require very small time steps.  This problem is not insurmountable, but one that may require significant additional computational time.  Overall, we estimate that we have successfully accounted for the following fractions of ULX-luminosity systems in Models A, B, C, and D: 66\%, 96\%, 92\%, and 90\%, respectively.  Thus, the appropriate corrections could be made for the corresponding ULX production rates that we present.

\section{Summary and Conclusions}
\label{sec:model}

\noindent
In this work we have explored a number of different models for ULX sources with intermediate-mass BH accretors.  These involved four different ranges of donor masses and initial orbital periods.  For each model, the evolution of a very large number of individual binary systems was computed. We find that in order to have a plausible efficiency for producing active ULX systems, there are two basic requirements for the capture of companion/donor stars.  First, the donor stars should be {\em massive}, i.e., $\gtrsim 8~M_\odot$.   Second, the initial orbital separations, after circularization, should be {\em close}, i.e., $\sim$$6-30$ times the radius of the donor star when on the ZAMS.  If these highly favorable conditions do occur, we see from Fig. 2 that up to $\sim$$25$\% of the IMBHs could have $L_x \gtrsim 10^{40}$ ergs s$^{-1}$, but only for a brief interval of time on the order of 1 Myr.  This is at least qualitatively consistent with the fact that such objects are quite rare, occurring, on average, only once in every $\sim$100 galaxies (see, e.g., Ptak \& Colbert 2004).

We make the following crude estimate of the number of ULXs with $L_x \gtrsim 10^{40}$ ergs s$^{-1}$ that might be found in a typical spiral galaxy:
\begin{equation}
N_{\rm IMBH-ULX} \simeq N_{\rm ysc} \cdot f_{\rm IMBH} \cdot f_{\rm cap} \cdot f_{\rm on} \cdot f_{\rm dur} ~~~,
\end{equation}
where $N_{\rm ysc}$ is the steady-state number of young (i.e., $\lesssim$50 Myr), massive star clusters potentially capable of producing an IMBH, $f_{\rm IMBH}$ is the fraction of all such clusters which produce an intermediate-mass BH, $f_{\rm cap}$ is the fraction of IMBHs that capture a massive star into a close orbit, $f_{\rm on}$ is the fraction of systems that would be ``on'' with this luminosity at a particular time (see Fig. 2), and $f_{\rm dur}$ is the fraction of the cluster lifetime when the massive stars would be transferring mass to the IMBH at a sufficiently high rate (see Fig. 2).   For illustrative parameter values we might take $N_{\rm ysc} \simeq 10$ (if anything, a generous estimate), $f_{\rm IMBH} \simeq 0.1$, $f_{\rm cap} \simeq 0.05$ (see Blecha et al. 2005), and $f_{\rm on} \times f_{\rm dur}$ ranges between 0.001 and 0.02 (as derived from Fig. 2).  We would then find $N_{\rm IMBH-ULX} \sim 5 \times 10^{-5}-10^{-3}$.  Thus, only one normal galaxy in $\sim 10^3-10^4$ might be expected to harbor a ULX with  $L_x \gtrsim 10^{40}$ ergs s$^{-1}$, about a factor of 10-100 lower than is indicated by the observations (see, e.g., Ptak \& Colbert 2004).  However, needless to say, a number of these parameter values are highly uncertain.

We have not considered in any depth either the formation of the IMBH itself, or the capture of the companion/donor star.  Nor have we taken into account the possibility that more than one companion star could be captured by the IMBH within the same time interval, and these could interact dynamically (see Pfahl 2005; Blecha et al. 2005).  However, at least for (i) modestly low cluster central star densities ($\sim 3 \times 10^4$ pc$^{-3}$), (ii) the duration of the evolutionary timescales of the massive donor stars of interest, and (iii) a reasonable binary fraction of $\sim$50\%, this may be a good approximation.  In summary, we conclude that in order for the IMBH model of ULXs to be viable, at a minimum, there must be a high efficiency for producing IMBHs in young star clusters, and a high probability of capturing a {\em massive} companion star into a relatively {\em close} orbit.  Even then, however, it is not obvious that the IMBH-ULX model can account for the observed systems.

Finally, even for our models with massive donor stars, IMBH systems with $10^{39} \lesssim L_x \lesssim 10^{40}$ ergs s$^{-1}$ are only $\sim$$1.5-2.5$ times more numerous than systems with $L_x \gtrsim 10^{40}$ ergs s$^{-1}$.  This ratio is inconsistent with the rapid falloff in ULXs with luminosity that is observed.  

\acknowledgements 

We thank Alan Levine for helpful discussions, and Robert Harris for his participation 
at an early stage of this study.
SR acknowledges support from NASA Chandra Grant NAG5-TM5-6003X.
LN thanks NSERC (Canada) and the CRC Program for financial support, and also
acknowledges the CCS at the Universite de Sherbrooke for their technical assistance.

\vspace{1 cm}


\begin{thebibliography}{}

\bibitem[]{} Alexander, T., \& Morris, M. 2003, ApJ, 590, 25.
\bibitem[]{} Bardeen, J.M. 1970, Nature, 226, 64.
\bibitem[]{} Baumgardt, H., Makino, J., \& Ebisuzaki, T.  2004, ApJ, in press [astro-ph/0406231].
\bibitem[]{} Begelman, M. 2002, ApJ, 568, 97.
\bibitem[]{} Blecha, L., Ivanova, N., Kalogera, V., Belczynski, K., Fregeau, J., \& Rasio, F. 2005, submitted to ApJ [astro-ph/0508597].
\bibitem[]{} Colbert, E., \& Mushotzky, R. 1999, ApJ, 519, 89.
\bibitem[]{} Colbert, E.J.M., \& Miller, M.C. 2004, talk at the Tenth
Marcel Grossmann Meeting on General Relativity, Rio de Janeiro, July
20-26, 2003. Proceedings edited by M. Novello, S. Perez-Bergliaffa and
R. Ruffini, World Scientific, Singapore, 2004 [astro-ph/0402677].
\bibitem[]{} Cropper, M. Soria, R., Mushotzky, R.F., Wu, K., Markwardt, C.B.,
 \& Pakull, M. 2004, MNRAS, 349, 39.
\bibitem[]{} Fabbiano, G. 1989, ARA\&A, 27, 87.
\bibitem[]{} Fabbiano, G., Zezas, A., \& Murray, S.S.  2001, ApJ, 554, 1035.
\bibitem[]{} Fabbiano, G., \& White, N.E.  2004, to appear in
``Compact Stellar X-Ray Sources'', eds. W.H.G. Lewin and M. van der
Klis, (Cambridge Univ. Press: Cambridge) [astro-ph/0307077].
\bibitem[G{\" u}rkan et al.(2004)]{2004ApJ...604..632G} G{\" u}rkan, M.~A., Freitag, M., \& Rasio, F.~A.\ 2004, \apj, 604, 632.
\bibitem[]{} Hopman, C., Portegies Zwart, S.F., \& Alexander, T. 2004, ApJ, 604, L101.
\bibitem[]{} Kalogera, V., Henninger, M., Ivanova, N., \& King, A.R. 2004, ApJ, 603, L41.
\bibitem[]{} King, A.R., Davies, M.B., Ward, M.J., Fabbiano, G., \&
Elvis, M. 2001, ApJ, 552, L109.
\bibitem[]{} K\"ording, E., Falcke, H., \& Markoff, S. 2002, A\&A, 382, L13.
\bibitem[]{} Miller, J.M., Fabbiano, G., Miller, M.C., \& Fabian,
A.C. 2003, ApJ, 585, L37.
\bibitem[]{} Miller, J.M., Fabian, A.C., \& Miller, M.C. 2004, ApJ, 607,
931.
\bibitem[]{} Nelson, C., \& Eggleton, P. P. 2001, ApJ, 552, 664.
\bibitem[]{} Pakull, M., \& Mirioni, L. 2003, RMxAC, 15, 197.
\bibitem[]{} Patruno, A., Colpi, M., Faulkner, A., \& Possenti, A. 2005, submitted to MNRAS [astro-ph/0507229].
\bibitem[Paxton (2004)]{Pax04} Paxton, B. 2004, PASP, 116, 699.
\bibitem[Pfahl(2005)]{2005ApJ...626..849P} Pfahl, E.\ 2005, \apj, 626, 849.
\bibitem[]{} Podsiadlowski, Ph. 1996, MNRAS, 279, 1104.
\bibitem[]{} Podsiadlowski, Ph., Rappaport, S., \& Pfahl, E. 2002, ApJ, 565, 1107.
\bibitem[]{} Podsiadlowski, Ph., Rappaport, S., \& Han, Z. 2003,
MNRAS, 341, 385.
\bibitem[Pols et al. (1995)]{Pols95} Pols, O.R, Tout, C.A, Eggleton, P.P \& Han, Zh.  1995, MNRAS, 274, 964.
\bibitem[]{} Portegies Zwart, S., \& McMillan, S.L.W. 2002, ApJ, 576, 899. 
\bibitem[]{} Portegies Zwart S.F., Dewi, J., \& Maccarone, T.  2004, MNRAS, 355, 413.
\bibitem[]{} Ptak, A., \& Colbert, E. 2004, ApJ, 606, 291.
\bibitem[]{} Rappaport, S., Podsiadlowski, Ph., \& Pfahl, E. 2005a, MNRAS, 356, 401.
\bibitem[]{} Rappaport, S., Podsiadlowski, Ph., \& Pfahl, E. 2005b, Proceedings of the Workshop on Interacting Binaries; Accretion, Evolution, \& Outcomes; Cefalu, Sicily, eds. L.A. Antonelli, et al., in press.
\bibitem[]{} Ray, A., Kembhavi, A. K., \& Antia, H. M. 1987, A\&A,184, 164.
\bibitem[]{} Roberts, T., \& Warwick, R. 2000, MNRAS, 315, 98.
\bibitem[]{} Ruszkowski, M., \& Begelman, M.C. 2003, ApJ, 586, 384.
\bibitem[]{} Wolter, A., \& Trinchieri, G. 2004, A\&A, 426, 787.
\end{thebibliography}
\end{document}